\documentclass[sigconf]{acmart}
\renewcommand\footnotetextcopyrightpermission[1]{} 

\usepackage{array}
\usepackage[ruled,linesnumbered]{algorithm2e}
\usepackage{amsfonts}
\usepackage{booktabs}
\usepackage{caption}
\usepackage{colortbl}
\usepackage{hhline}
\usepackage{makecell}
\usepackage{mathtools}
\usepackage{marginnote}
\usepackage{microtype}
\usepackage{multirow}
\usepackage{setspace}
\usepackage{subcaption}
\usepackage{tabularx}
\usepackage{tabulary}
\usepackage{xargs}
\usepackage{xcolor}
\usepackage{hyperref}

\everypar{\looseness=-1}
\makeatletter
\usepackage{etoolbox}
\apptocmd{\sloppy}{\hbadness 10000\relax}{}{}
\newcommand{\results}[1]{95\%}
\newcommand{\vect}[1]{\lowercase\textbf{\textit{#1}}}
\newcommand{\matrx}[1]{\uppercase\textbf{\textit{#1}}}
\def\mca#1{\multicolumn{1}{c}{#1}}
\def\mcb#1{\multicolumn{1}{c|}{#1}}
\def\mcr#1{\multicolumn{1}{r}{#1}}
\newcommand{\jsma}{\textsc{JSMA}}
\newcommand{\ajsma}{\textsc{AJSMA}}

\newcommand{\hsg}{\textsc{HSG}}

\newcommand{\cc}[1]
{
  \ifdim#1pt>45.0pt
    {\cellcolor{black!#1}{\textcolor{white}{#1\%}}}
  \else
    {\cellcolor{black!#1}{#1\%}}
  \fi
}

\begin{document}
\date{}
\title{Adversarial Examples in Constrained Domains}
\author{Ryan Sheatsley}
\email{sheatsley@psu.edu}
\author{Patrick McDaniel}
\email{mcdaniel@cse.psu.edu}
\affiliation{%
  \institution{The Pennsylvania State University}
}

\author{Nicolas Papernot}
\authornote{Work done while the author was at the Pennsylvania
State University}
\authornotemark[0]
\email{papernot@google.com}
\affiliation{%
\institution{Google Brain}
}

\author{Michael J. Weisman}
\email{michael.j.weisman2.civ@mail.mil}
\author{Gunjan Verma}
\email{gunjan.verma.civ@mail.mil}
\affiliation{%
\institution{United States Army Research Laboratory}
}



\begin{abstract}

\everypar{\looseness=-1}\noindent Machine learning algorithms have been shown to be
vulnerable to adversarial manipulation through systematic modification of inputs
(e.g., adversarial examples) in domains such as image recognition. In the default
threat model, the adversary exploits the unconstrained nature of images--each
feature (pixel) is fully under control of the adversary. However, it is not clear
how these attacks translate to \textit{constrained domains} (e.g., network
intrusion detection) that limit which and how features can be modified by the
adversary. In this paper, we explore whether constrained domains are less
vulnerable than unconstrained domains to adversarial example generation
algorithms. We create an algorithm for generating \textit{adversarial sketches}:
targeted universal perturbation vectors that encode feature saliency within the
envelope of domain constraints. To assess how these algorithms perform, we
evaluate them in constrained (e.g., network intrusion detection) and unconstrained
(e.g., image recognition) domains. The results demonstrate that our approaches
generate misclassification rates in constrained domains that were comparable to
those of unconstrained domains (greater than \results{}). Our investigation shows
that the narrow attack surface exposed by constrained domains is still
sufficiently large to craft successful adversarial examples--and thus, constraints
do not appear to make a domain robust.  Indeed, even if a defender constrains an
adversary to as little as five random features, generating adversarial examples is
still possible.
\end{abstract}

\setcopyright{acmcopyright}
\copyrightyear{2018}
\acmYear{2018}
\acmDOI{10.1145/1122445.1122456}

\acmConference[London '19]{London '19: ACM Symposium on Computer and 
  Communications Security}{November 11--15, 2019}{London, UK}
\acmBooktitle{London '19: ACM Symposium on Computer and 
  Communications Security,
  November 11--15, 2018, London, UK}
\acmPrice{15.00}
\acmISBN{978-1-4503-9999-9/18/06}


\maketitle
\pagestyle{plain} 

\section{Introduction} \label{sec:introduction} \looseness=-1 
Machine learning algorithms are
rapidly revolutionizing many industries including
transportation ~\cite{pl91efficient}, finance ~\cite{bs01business,wg97neural},
health care ~\cite{kr15machine}, education~\cite{jl08authorship,sz07machine}, and
security ~\cite{db15aso,sm10outside,sc99application,ts09intrusion}. For the past
decade, we have seen a revolution in automation as research has focused on
increasing the accuracy, problem size, and applicable domains for these automated
learners. The results are promising: the latest learning algorithms have shown
previously impossible accuracies for problems spanning multiple domains. However,
when an adversary is introduced, the machine learning algorithms and the myriad
domains they serve often become vulnerable to an adversary.

\looseness=-1
One of the directions the field of adversarial machine learning explores is the
impact of \textit{adversarial examples}: inputs to machine learning models that an
attacker has intentionally designed to cause the model to make a mistake, e.g.,
misclassify~\cite{gf17attacking}. Within the scope of this research, there have
been studies that target particular inputs through a variety of threat models,
with algorithms such as the \textsc{Jacobian-Saliency Map
Approach}~\cite{pp16limit}, \textsc{Carlini-Wagner}~\cite{cl17towards}, and
\textsc{Projected Gradient Descent}~\cite{md18towards}, among others. A recent
class of attacks, exposing potentially more serious vulnerability, compute
\textit{universal adversarial perturbations}~\cite{gf15explain,df16universal}.
These perturbations can be precomputed and quickly applied at runtime to arbitrary
inputs and still achieve adversarial goals (e.g., misclassification of arbitrary
inputs to a class chosen by the adversary). The existence of adversarial examples
presents a compelling barrier for sensitive domains that use machine learning.
Investigations have shown that no domain (thus far) is immune to this phenomenon;
the scope of adversarial examples has been expansive, reaching into image
processing~\cite{pp16limit,gf15explain,cl17towards,ms16deep}, malware
detection~\cite{gs17adversarial,kj18adversarial},
text~\cite{ji17adversarial,eh18hotflip}, and even speech
recognition~\cite{cl18audio}.

\looseness=-1
A repeated criticism of adversarial machine learning research is that
investigations have almost completely focused on unconstrained domains. 
Implicit to this argument is that such freedom overestimates the capabilities of an
adversary in constrained domains where they are often bound by the semantics
of features \textit{and} capable of only controlling a subset of features. It stands to
reason that such algorithms used to generate adversarial examples would be less
effective in constrained domains. In this work, we consider network intrusion
detection as our exemplary constrained domain, using existing network intrusion
detection datasets: the NSL-KDD~\cite{tl09detailed} and UNSW-NB15~\cite{ms15unsw}.
Here, \textit{constraints} are defined by the following three characteristics: the
values within a feature may be fixed (binary vs continuous), the values of
different features may be correlated (TCP flags in packets and TCP as the
transport protocol), and some features may not be controllable by an adversary
(round-trip times).

\looseness=-1
Furthermore, while there has been research that has focused on constrained 
\textit{adversaries}\cite{gs17adversarial,su19one}, who are bound in their 
capabilities (e.g., how many total features can be perturbed or which features 
are manipulable), we differentiate in that we consider constrained \textit{domains}.
Specifically, these constraints describe the kinds of inputs that are 
\textit{permissible} in a domain (For example, network packets that do not 
obey the TCP/IP protocol would not be permissible). In this work, we consider 
the union of both adversary and domain constraints, unlike previous work. 

\looseness=-1
In this paper, we test the hypothesis that constrained domains are less vulnerable
to the techniques of adversarial machine learning from two perspectives: from the
perspective of traditional adversarial algorithms, and from universal adversarial
perturbations. This leads to two approaches in our evaluation. Prior to our
evaluation, we identify and extract constraints from a dataset. Then, we develop
an augmented algorithm, the \textsc{Adaptive JSMA} (\ajsma{}), to construct
adversarial examples that obey domain constraints. Next, we design a second
algorithm, the \textsc{Histogram Sketch Generation} (\hsg{}), the first attack to
compute \textit{adversarial sketches}: universal perturbations used to craft
adversarial examples en masse that comply with domain constraints. Other
algorithms that produce universal perturbations neither have a mechanism to comply
with constraints, nor are they immediately suitable for attacking constrained
domains (as described in \S\ref{sec:methodology}). With both algorithms, we
measure the success rate of crafting adversarial examples by attacking models
directly in both constrained and unconstrained domains, and find that we were
comparably successful in both domains (greater than 95\%). Furthermore, we use
adversarial examples crafted from a model to attack different learning techniques
trained on similar, but not identical, data, demonstrating that adversarial
examples can \textit{transfer} in constrained domains (reaching up to 93\% for the
\ajsma{} and 100\% for the \hsg{}). Finally, we show that even if an adversary
maintains attack behavior \textit{and} cannot arbitrarily control certain features
\textit{and} must obey the TCP/IP protocol, there is still a surprising amount of
exploitable attack surface to craft adversarial examples. We make four
contributions:

\begin{enumerate}

  \item We introduce formalism to express constraints and design an algorithm that
    is able to extract them systematically in the domains we consider. Learning
    constraints codifies the space of permissible adversarial examples.

  \item We introduce two algorithms. The \textsc{Adaptive JSMA}, which crafts
    adversarial examples that obey domain constraints. The \textsc{Histogram
    Sketch Generation}, which produces adversarial sketches: universal adversarial
    perturbations that obey domain constraints.

  \item We perform experiments where a we impose an extreme amount of
    constraints, to the point where an adversary can only control five random
    features, and show that adversarial examples can still be crafted with a
    $\mathtt{\sim}50\%$ success rate within our studied constrained domains. 

  \item We demonstrate promising results for both algorithms, reaching greater
    than \results{} misclassification rates across the datasets used in our
    experiments. This suggests that our studied constrained domains are as
    vulnerable as their unconstrained counterparts.

\end{enumerate}
 
\section{Background} \label{sec:background} Adversarial machine learning research
in unconstrained domains has been broad. Since the initial observations of Biggio
et al. and Szegedy et al. in deep neural
networks~\cite{bg13evasion,sg13intriguing} to the robust attacks from Kurakin et
al. and Sharif et al.~\cite{kk17physical,sr16accessorize}, adversarial examples
have matured from, ``an intriguing property'' to a tangible threat.

\everypar{\looseness=-1}
The first generation of attacks were formed in the context of ``white-box''
attacks~\cite{pp16limit,gf15explain,cl17towards}. Under this threat model,
adversarial examples are crafted using information (e.g., model parameters)
directly from the model under attack. This represents a worst-case scenario,
analogous to an insider threat, since such information would not be easily
accessible in most practical contexts. Naturally, this motivates the question: Can
an adversary successfully attack a model, \textit{without having direct access to
its parameters?} Papernot et al. and Tram{\`e}r et al. investigated this question
by leveraging \textit{transferability}: an adversarial example crafted from one
model will often be an adversarial example for a different model, even if they are
using different training data and/or learning
techniques~\cite{tm17space,pp16transferability}. Through this ``black-box,''
threat model, an adversary trains a \textit{surrogate} model by using inputs to
generate output labels from the victim model, called an \textit{oracle}.
Afterwards, the surrogate model is used to craft adversarial examples which are
then (with high probability) ``transferred'' to the victim model.

Concurrently, others have investigated what limitations (if any) exist for
adversarial examples. Kurakin et al. and Brown et al. explored how adversarial
examples can be applied directly to the physical domain, introducing techniques
that produce adversarial examples robust to physical distortions, such as
rotation, scale, and other transformations~\cite{kk17physical,bn17adversarial}.
Moreover, Goodfellow et al. and Moosavi-Dezfooli et al. analyzed \textit{universal
adversarial perturbations}: single perturbation vectors used to quickly craft
adversarial examples from many inputs not known in
advance~\cite{gf15explain,df16universal}.  These universal perturbations are
particularly concerning as they enable adversaries to take computation offline and
amortize computational costs over many inputs. At present, we are observing an
evolution in the way adversarial examples manifest. Each generation contributes to
a growing threat against deployed machine learning systems.

For this work, we modify an existing attack, the \textit{Jacobian-based Saliency
Map Approach}, introduced by Papernot et al.~\cite{pp16limit}. The
``\textsc{JSMA}'' produces an adversarial example by iteratively applying
perturbations to the most salient features in an input. The algorithm terminates
when either the input is successfully misclassified or the specified $l_0$
distance\footnotemark is reached. The \jsma{} greedily selects features to perturb
by constructing \textit{saliency maps}, which encode the influence features have
over misclassifying a particular input. This $l_0$ minimization makes the \jsma{}
an attractive candidate for our evaluated domain, as discussed in
\S\ref{sec:methodology}. However, the \jsma{} is not necessary for crafting
adversarial examples in constrained domains. Different attack algorithms can be
used, with some adjustments, as reviewed in \S\ref{sec:methodology}, which we
defer to future work. For our study in constrained domains, the \jsma{} was simply
the most readily usable algorithm to leverage.

\footnotetext{\looseness=-1 Most attack algorithms have upper limits on the allowable distortion
they can introduce. This distortion is defined to be the distance between an
adversarial example and its original counterpart. There are many metrics
(principally $l_p$ norms~\cite{cl17towards}) used throughout the literature to
measure this distance.}

\section{Methodology} \label{sec:methodology} In this section, we explain the
intuition behind generating adversarial examples in constrained domains.

\subsection{AML in Constrained Domains} Throughout, both in this section for
examples and in the evaluation, we will focus principally on network intrusion
detection data (specifically, the use of TCP/IP) due to the constrained nature of
the domain. Here, we assume a network intrusion detection system classifies
feature vectors (representing network traffic flows) as benign or
malicious\footnotemark. These feature vectors are created through a feature
extraction algorithm, which accepts network packets as input.  Therefore, the
adversary seeks to bypass the network intrusion detection system by tweaking
malicious network traffic (guided by adversarial machine learning techniques), so
that it is subsequently classified as benign. Naturally, the adversary must obey
the TCP/IP protocol so that the network attacks can be realizable. Any feature
vector that violates the TCP/IP protocol (such as negative port numbers) would be
manifestly adversarial. 

\footnotetext{\looseness=-1 Without loss of generality, the models we use in the
evaluation discriminate between different kinds of network attacks (as well as
benign traffic). Nonetheless, the goal of the adversary remains unchanged.}

\subsection{Challenges in AML with Constraints} Crafting adversarial examples in
constrained domains is a necessarily different process from crafting adversarial
examples in unconstrained domains. Not all features represent the same kind of
information (pixels vs packet information), nor do they describe the same kind of
statistical data (discrete vs a blend of categorical, continuous, and discrete).
These differences change the threat surface and the underlying assumptions
surrounding the capabilities of an adversary in constrained domains. Existing
algorithms are unsuitable for attacking constrained domains for these reasons:

\looseness=-1
\vspace{2pt}\noindent\emph{(1) Existing algorithms are largely optimized for human
perception.} While there is an open discussion on the amount and kinds of
distortion that are appropriate~\cite{cl17towards}, existing algorithms have been
tuned for image domains~\cite{gf15explain,cl17towards,ms16deep}. That is, these
algorithms try to minimize human perception of the distortion introduced in
adversarial examples. However, such metrics have no meaning for many
constrained domains because they are not perceived by humans. Thus, using
algorithms optimized for human perception offers us limited utility.

\looseness=-1
\vspace{2pt}\noindent\emph{(2) Existing algorithms assume adversaries have full
control over the feature space.} Most algorithms perturb the entire feature space
to minimize an $l_p$ norm as a surrogate for estimating a measure of human
perception~\cite{gf15explain,cl17towards,ms16deep}. This is likely an unreasonable
assumption in constrained domains. For example, in network intrusion detection,
features can represent broad network behaviors that exist outside the control of
an adversary, e.g., round-trip times.

\vspace{2pt}\noindent\emph{(3) Existing algorithms do not consider domain
constraints.} Crafting adversarial examples that obey domain constraints is
necessary to mount practical attacks. However, prior works strategically applied
perturbations in order to avoid constraints~\cite{gs17adversarial,kj18adversarial}
(such as only adding bytes at the end of a binary). While a reasonable strategy
for the studied domain (malware), it may not be effective if the adversary can
only perturb features that are constrained. As an example of a constraint, network
intrusion detection datasets commonly have protocol and service (port number) as
features and certain services are exclusive to certain protocols. Therefore, to
produce an adversarial example that is representative of a legitimate traffic
flow, algorithms need to enforce these constraints.

\subsection{Crafting with Constraints} In this subsection, we describe how we
learn constraints and incorporate them into crafting adversarial examples.

\looseness=-1 
\vspace{3pt}\noindent\textbf{Learning Constraints.} If the constraints are not
explicitly given, then they must be inferred. One of the first requirements for
properly modeling the constraints of any modeled domain is to codify the specific
relationships between features. One of the most important of these relationships
is the notion that we introduce as a \textit{primary feature.} Intuitively, a
primary feature is a feature that, when set to particular value, limits the range
of permissible values for other features. Conversely, a \textit{secondary feature}
does not impose any limitations for other features.  Therefore, the first task is
to identify primary features and the relationships from these primary features to
secondary features. Interestingly, primary features may have relationships amongst
themselves, as well as secondary features.

\looseness=-1 As an example, many popular network intrusion detection datasets
include features that represent services, packet flags, and other protocol-related
information. Since these features share a causal relation with protocols, we
designate transport layer protocols to be primary features and the others as
secondary features\footnotemark. As another example, we could designate using SSL
as a primary feature, which would imply using TCP as the transport protocol, which
finally dictates the values for other secondary features. Table
\ref{tab:nsl-constraints} in the \hyperref[sec:appendix]{Appendix} demonstrates
the constraints for one of our evaluated datasets.

\footnotetext{\looseness=-1 We identified our primary features based on our
understanding of the domain and observations in the data. We explore methods to
systematically discover primary features when domain expertise is unavailable in
\S\ref{sec:discussion}.}

\looseness=-1
Given that constraints are restrictions on where and how features can be
perturbed, it is intuitive to model these constraints as a simple form of
first-order logic. Here, primary features are the predicates in logical
expressions that determine the properties (e.g., values) that a collection of
variables (e.g., secondary features) can have. Said alternatively, the set of
primary features are the conditionals on the values of other secondary features.
Formally, we can understand constraints to have the following form:
\begin{equation*} \forall \vect{x} \in \mathbb{X}: \vect{x}_k \Rightarrow
(\vect{x}_1 \in \mathbb{Y}_1) \land (\vect{x}_2 \in \mathbb{Y}_2) \land \dots
\land (\vect{x}_n \in \mathbb{Y}_n) \end{equation*} 
\noindent where $\vect{x}$ represents an input in a dataset $\mathbb{X}$, $k$ is a
primary feature, and $\mathbb{Y}_n$ represents the values permissible by the
semantics of feature $n$ (e.g., $\{0,1\} \in \mathbb{Y}$ if the feature is binary,
or $\{\mathbb{R}: 0 \leq y \leq 1\} \in \mathbb{Y}$ if the feature is continuous).
For example, we can represent a simple TCP constraint as follows:
\begin{equation*}
  \forall \vect{x} \in \mathbb{X}: \vect{x}_{TCP} \Rightarrow \vect{x}_{port} \in
  [1,\dots, 65535]
\end{equation*}
\noindent There are many existing algorithms that can learn constraints from
logic~\cite{dr18learning}, as inputs can be modeled as simple instantiations of
expressions (i.e., \textit{propositions}). Here, we use primary features to design
an algorithm to learn these constraints.  

After primary features have been identified, we begin to learn constraints based
on the following heuristic: a constraint exists between a primary feature $k$, and
any other feature $p$, if there exists at least one input in the training set
where both $k$ and $p$ are seen together. For example, features that describe TCP
packet flags (i.e., $p$) would have the value 1.0 for TCP traffic flows (i.e.,
$k$) and 0.0 for non-TCP traffic flows. Therefore, TCP packet flag features are
\textit{constrained} to TCP flows. Conceptually, these constraints encode the
maneuvers that are possible (and probable) for an adversary\footnotemark.

\footnotetext{In \S\ref{sec:uncontrollable}, we also perform experiments where we
further add additional constraints that an adversary would have to obey.}

\looseness=-1 \vspace{3pt}\noindent\textbf{Addressing Constraints.} To address the
challenges discussed prior, we integrate \textit{constraint resolution} into the
crafting process. This guarantees that generating an adversarial example obeys not
only the semantic constraints of the domain, but also the probabilistic
constraints of the dataset as well. As an artifact of our constraint learning
process, we may also learn constraints that are technically allowable within the
semantics of the domain, but are simply never observed in the dataset. As a second
technique, we also simultaneously minimize the total number of features perturbed
(i.e., $l_0$) to obey constraints and comply with the limited control over
features an adversary may have. 

\vspace{3pt}\noindent\textbf{Measuring Distance.} The distance between an
adversarial example and its original counterpart is a measurement of the
distortion introduced by an algorithm, commonly represented by an $l_p$ norm. Most
algorithms have either limits on the maximum allowable distortion or explicit
termination conditions when a particular amount of distortion is introduced. There
is debate on the most appropriate distance measure (i.e., the choice of $l_p$
norm) to use for modeling levels of human perception~\cite{cl17towards}. Our study
of constrained domains departs from this debate, as most of these domains are not
inherently visual. Therefore, we argue that measuring distance under the $l_0$
norm is appropriate for our purposes. In particular, we choose $l_0$ to model how
an adversary may be limited to controlling certain features, to enforce domain
constraints, and because we are not optimizing for human perception.

\looseness=-1
\vspace{3pt}\noindent\textbf{Augmenting the JSMA.} Before we can use the \jsma{}
in constrained domains, we make a slight adjustment to the algorithm so that it
can better search the space of possible adversarial examples. By default, the
\jsma{} parameter $\theta$ determines the magnitude and direction of a selected
perturbation. A positive $\theta$ will increase feature values and a negative
$\theta$ will decrease them. To allow an adversary more freedom when crafting
adversarial examples, we modified the \jsma{} to perturb in either direction
dynamically. This directly improved the success rate of the \jsma{}. We refer to
this modified version as the \textsc{A(daptive)} \jsma{}. 

To allow the \ajsma{} to perturb in either direction, we simply evaluate both masks
used in~\cite{pp16limit} that are applied to the saliency map when $\theta$ is
positive or negative. Formally, for any feature $i$ to be a perturbation
candidate, $i$ must satisfy:

\footnotesize
\[
  \begin{split}
  \left(\frac{\partial f_t(\vect{x})}{\partial \vect{x}_i} > 0 \text{ and }
  \sum_{j\neq t} \frac{\partial f_j(\vect{x})}{\partial \vect{x}_i} < 0\right)
  \text{ or } \left(\frac{\partial f_t(\vect{x})}{\partial \vect{x}_i} < 0
  \text{ and } \sum_{j\neq t} \frac{\partial f_t(\vect{x})}{\partial
  \vect{x}_i} > 0\right)
  \end{split}
\]
\normalsize

\noindent where $\frac{\partial f_t(\vect{x})}{\partial \vect{x}_i}$ represents
the forward derivative for a model $f$ and target class $t$ with respect to
feature $i$ in an input $\vect{x}$. Conceptually, the \ajsma{} only considers
features to be perturbable if the target gradient and sum of non-target gradients
are in opposing directions. Intuitively, this simply means that any perturbation
reduces the distance to the target class or increases the distance to non-target
classes.

This modification enables us to determine the optimal perturbation direction for
increasing (the left half of the mask) or decreasing (the right half of the mask)
features. Afterwards, we use the scoring metric found in~\cite{pp16limit} and
return the most influential feature with the optimal perturbation direction. 

\looseness=-1
\vspace{3pt}\noindent\textbf{Integrating Constraints into the \ajsma{}.}
Integrating constraint resolution distills to preventing the \ajsma{} from
selecting  features that violate constraints. Once a  feature is
selected for perturbation, we check if this feature is constrained to a primary
feature\footnotemark. This check is described by Algorithm \ref{alg:constraints}.

\footnotetext{We note that any additional perturbations made by our constraint 
resolution algorithm are included when we measure the distance between an
original input and an adversarial example. To this effect, adversarial examples
that obey domain constraints do not violate $l_0$ norms.}

Consider the following example using the Algorithm \ref{alg:constraints}. A UDP
traffic flow is given to the \ajsma{}. After analyzing the saliency map, the
\ajsma{} suggests that the current service, \textsc{tftp\_u}, should be switched
to \textsc{ftp}, which is a service constrained to TCP. After the service switch
is made, the perturbed input is presented to Algorithm \ref{alg:constraints} to
check whether or not any constraint is violated. The first condition determines if
the perturbed feature $p$ is a primary feature (i.e., TCP, UDP, or ICMP). In this
example, it is not, so we move to the second condition and evaluate if $p$ is
constrained to exclusively one primary feature. In this case, $p$ is exclusively
associated with TCP. Thus, the search domain is further restricted to
TCP-compliant features and the transport protocol of the input is switched from
UDP to TCP. Since the input has switched primary features from UDP to TCP, we
finally set all non-TCP features to 0. Once the \ajsma{} terminates, the produced
adversarial example will be representative of a permissible traffic flow, i.e., it
obeys the domain constraints.

\begin{spacing}{0.1}
\begin{algorithm}[!t]

  \footnotesize
  \DontPrintSemicolon
  \SetAlgoNoLine
  \SetKwInOut{Input}{Input}
  \Input{$p$, $\Gamma$, $\matrx{S}$, $\vect{X}$, $h$}
  \tcp{p is a primary feature}
  \If{$p\in \mathbb{K}$} {
    $\Gamma = \Gamma\cap h(p)$ \\
    $\vect{x}_p\leftarrow$ switch primary feature to $p$\\
  }
  \tcp{p is constrained to exactly one primary feature}
  \ElseIf{$\exists! k\in \mathbb{K} \textrm{ \upshape s.t. } p\in h(k)$}{
    $\Gamma = \Gamma\cap h(k)\setminus\{p\}$\\
    $\vect{x}_k\leftarrow$ switch primary feature to $k$\\
  }
  \tcp{p is constrained to multiple primary features}
  \ElseIf{$\exists k\in \mathbb{K} \textrm{ \upshape where } p\in h_k$}{
    $\Gamma = \Gamma\setminus\{p\}$\\
    $k'\leftarrow$ the current primary feature in $\vect{x}$\\
    \tcp{x is using an illegal primary feature wrt p}
    \If{$p\notin h(k')$}
    {
      $k= \textrm{arg max}_{\{k\in\mathbb{K} \rvert p\in h(k)\}} \, S_k$\\
      $\Gamma = \Gamma\cap h(k)$\\
      $\vect{x}_k\leftarrow$ switch primary feature to $k$\\
    }
  }
  \tcp{p constrains all primary features}
  \ElseIf{$\forall k \in \mathbb{K}, p\in h_k$}{
    $\Gamma = \Gamma\setminus\{p\}$\
  }
  \If{\textrm{ \upshape switched primary feature to }$k$}{
    \tcp{ensure x is not using illegal features}
    $\forall i \notin h(k),\, \vect{x}_i\leftarrow 0$
  }
  \textbf{return} $\Gamma$, $\vect{x}$
  \caption{\textsc{Resolving Constraints\newline}
  \label{alg:constraints}
  \footnotesize
  \textrm{$p$} is a candidate feature, \textrm{$\Gamma$} is the search domain,
  $\matrx{s}$ is the saliency map for the current input $\vect{x}$,
  $h:\mathbb{K}\mapsto\mathbb{V}$ is an associative array containing
  constraints.}

\end{algorithm}
\end{spacing}

\subsection{Creating Adversarial Sketches} In this subsection, we describe how we
create adversarial sketches\footnotemark: universal 
perturbations that obey constraints.

\footnotetext{\looseness=-1 The concept of ``sketching,'' also known as
\textit{approximate query processing}~\cite{cm11sketch}, was first introduced by
Flajolet et al.  Sketching refers to a class of streaming algorithms that seek to
extract information from a data stream in a single pass~\cite{fj85probabilistic}.
Commonly deployed in memory-constrained environments, these algorithms approximate
or summarize the information in a given data stream. Adversarial sketches are
similar as they are an approximation of a universal perturbation and computed
through one pass of inputs.}

For the same reasons described at the beginning of \S\ref{sec:methodology},
existing algorithms to generate universal adversarial
perturbations~\cite{hy17service,df16universal,bn17adversarial} are not immediately
usable in constrained domains: the algorithms are largely optimized for human
perception, they assume adversaries have full control over the feature space, and
they do not consider domain constraints.

\looseness=-1
In our search for adversarial sketches in constrained domains, we take a
principled approach using two tools: adversarial examples generated from the
\ajsma{} and a \textit{perturbation histogram}. Prior to the algorithms mentioned
earlier for computing universal adversarial perturbations, encountering these
perturbations was a matter of chance: an adversary would be required to apply a
perturbation generated from an attack on other unperturbed inputs and simply
observe the universality of the perturbation, i.e., brute-forcing the universal
perturbation~\cite{gf15explain}.

\looseness=-1
\vspace{3pt}\noindent\textbf{Adversarial Examples from the AJSMA.} Initially, we
followed the same brute-force approach in~\cite{gf15explain}: we used adversarial
examples crafted from the \ajsma{} to glean insights for universal perturbations.
This brute-force approach is feasible (in terms of computational complexity) for
network intrusion detection datasets as their dimensionality and cardinality is
small. After evaluating every perturbation generated by the \ajsma{} on every
input in the test set, we discovered a handful of universal perturbations.

\looseness=-1
\vspace{3pt}\noindent\textbf{Perturbation Histograms.} The perturbation histogram
encodes how perturbations are distributed en masse (an example is shown later in
\S\ref{sec:evaluation}). To produce the histogram, we enumerate over all of the
perturbations generated by the \ajsma{} (including any additional perturbations
made to resolve constraints) and record the perturbed features and directions. Our
insight for the histogram is rooted in how the \ajsma{} scoring metric ranks
influential features; we hypothesized that features commonly perturbed across
inputs from different classes would be optimal candidates for building an
adversarial sketch. This hypothesis was reinforced by our observation that the
perturbation histogram was (relatively) static: random shuffling of partitioned
training sets, unique training parameters, and random subsets of analyzed
adversarial examples yielded minor changes to the perturbation histogram.  These
substantial adjustments to our experiment workflow demonstrated little change
between perturbation histograms. These observations suggest that the perturbation
histogram is a combined representation of class-based saliency \textit{and} domain
constraints.

\looseness=-1
With the universal perturbations discovered through the \ajsma{} and the static
nature of the perturbation histogram, we made an observation: \textbf{the majority
of the perturbed features (and their associated directions) in the most successful
universal perturbations generated from the \ajsma{} mapped \textit{directly} to
the most perturbed features in the perturbation histogram. This key observation
led us to the creation of the \textit{Histogram Sketch Generation}}.
 
\looseness=-1
\vspace{3pt}\noindent\textbf{Histogram Sketch Generation.} The \textsc{Histogram
Sketch Generation} accepts a perturbation histogram $\matrx{h}$ and integer $n$ as
parameters and returns a adversarial sketch $\vect{a}$, which consists of the top
$n$ most frequently perturbed features and optimal directions\footnotemark. We
observed that the most successful universal perturbations generated by the
\ajsma{} had a subset of features that directly mapped to the most frequently
perturbed features in the perturbation histogram.  Intuitively, if we consider
these successful universal perturbations as the optimal solution, then the \hsg{}
approximates the optimal solution by returning a subset of those features (and
associated directions). Figure \ref{fig:graphics} in the
\hyperref[sec:appendix]{Appendix} demonstrates a handful of sketches.

\footnotetext{While the \hsg{} does not take a target class as a parameter, it
uses information directly from the perturbation histogram to create an adversarial
sketch. As a consequence, the \hsg{} will return a targeted adversarial sketch if
the histogram is built from targeted adversarial examples.}

\looseness=-1
It is interesting to note that the \hsg{} is essentially the problem of variable
selection in classical statistics. Variable selection involves the selection of a
subset of relevant variables (or features) in the model, such that a ``minimal''
amount of information is lost (e.g., minimal impact on model loss). Many common
procedures in classical statistics, such as LASSO or stepwise regression, aim to
balance model fit with a penalty on the $l_0$ norm (or a relaxation of this norm
to $l_1$, in the case of LASSO)\cite{tr96lasso}. Our approach of greedily
selecting the top $n$ most perturbed features is directly analogous to stepwise
regression's (greedy) selection of the $n$ features which best explain the data.
As the complexity grows combinatorially with the number of features, greedy
methods are necessarily resorted to, and our approach here is no exception.

\section{Evaluation} \label{sec:evaluation} In this section, we evaluate our
approach on two network intrusion datasets, the NSL-KDD and the UNSW-NB15, as
well as two image recognition datasets, the GTSRB and MNIST. 
In our evaluation, we answer two questions:

\begin{enumerate}

  \item Are constrained domains more robust against adversarial examples?
  \item Do universal adversarial perturbations exist in constrained domains?

\end{enumerate}

Our experiments revealed that: we can craft adversarial examples with success
rates greater than \results{}, even in the presence of constraints; we can compute
highly successful adversarial sketches, reaching greater than 80\%
misclassification rates for the majority of learning techniques.

Our experiments were performed on a Dell Precision T7600 with an Intel Xeon
E5-2630 and a NVIDIA GeForce TITAN X. We used Cleverhans
2.0.0~\cite{pp16cleverhans} for training our models and crafting adversarial
examples. 

\subsection{Datasets}\looseness=-1 Before we describe our experiments, we provide
an overview of the four evaluated datasets and any preprocessing that we
performed. Table \ref{tab:architecture} describes the model architectures,
hyperparameters, and model accuracies across all four datasets\footnotemark. 

\footnotetext{Note that while
the accuracy of our network intrusion detection models sit around $77\%$, these
are consistent with the literature~\cite{tl09detailed,
ig15performance,ih13comparison,ms17hybrid}. Notably, Tavallaee et al., who 
initially uncovered the inherent problems with the KDD CUP 99 (which 
resulted in the creation of the NSL-KDD), reported 77.41\% accuracy with 
Multi-layer Perceptrons~\cite{tl09detailed}. Furthermore, Moustafa et al., who 
designed the UNSW-NB15 to address the shortcomings of the 
NSL-KDD~\cite{ms15unsw}, reported 81.34\% accuracy (using the full training 
set) with an artificial neural network~\cite{mt16evaluation}. These accuracies 
are are comparable to the highest to date.}

We evaluate our approach on two network intrusion detection datasets for the
application of constrained domains and two image classification datasets for
unconstrained domains. We use these two image classification datasets for
comparison with other works in adversarial machine learning as well as a
demonstration of the cross-domain applicability of our approach. Note that these
are unconstrained domains (i.e., images), and thus the \ajsma{} behaves similarly
to the original \jsma{} (other than that it can perturb in either direction).

\begin{table}[!ht]
  \resizebox{\columnwidth}{!}{ \centering 
  \begin{tabular}{lcccccr}
    \toprule

    {Dataset} & {Architecture} & {Units} & {Batch Size} & {Learning Rate} & {Epochs} & {Testing Acc.}\\
    \midrule
    {\textbf{NSL-KDD}} & {MLP} & {123, 64, 32, 5} & {200} & {0.01} & {5} & {$77\%\pm 1.0\%$}\\
    {\textbf{UNSW-NB15}} & {MLP} & {196, 98, 49, 10} & {128} & {0.01} & {10} & {$75\%\pm 1.2\%$}\\
    {\textbf{MNIST}} & {CNN} & {784, 128, 128, 10} & {128} & {0.001} & {6} & {$98\%\pm 0.1\%$}\\
    {\textbf{GTSRB}} & {CNN} & {2700, 128, 128, 42} & {128} & {0.001} & {6} & {$82\%\pm 2.0\%$}\\

    \bottomrule
  \end{tabular}}
  \caption{Model Information}
  \label{tab:architecture}

\end{table}

\looseness=-1
\vspace{3pt}\noindent\textbf{NSL-KDD.} The NSL-KDD dataset is an improved variant
of the KDD Cup99 dataset~\cite{tl09detailed}. The KDD Cup99 (and its NSL-KDD
successor) have been used widely in the network intrusion detection community. We
chose to use the NSL-KDD for the novel application of adversarial examples in this
field, familiarity of the dataset within the academic community, and the lack of
well-formed network intrusion detection data. 

\looseness=-1
The NSL-KDD contains 5 classes, with 4 attack classes and 1 benign class.  It
contains 125,973 samples for training and 22,543 samples for testing.  It contains
41 features\footnotemark, separated into four high-level categories of features:
basic features of TCP connections, content features within a connection suggested
by domain knowledge, traffic features that are computed using a two-second time
window, and host-based features. The NSL-KDD has been widely studied and so we
defer to prior work~\cite{tl09detailed, db15aso} for the subtle details of the
dataset.

\footnotetext{We used the post-processing options in WEKA~\cite{hl09weka}, an
open-source data mining framework, to convert categorical features to one-hot
vectors.}

\looseness=-1 \vspace{3pt}\noindent\textbf{UNSW-NB15.} The UNSW-NB15 dataset was
designed to be an updated version of the NSL-KDD, containing modern attacks that
express a ``low footprint''~\cite{ms15unsw}. The Australian Centre for Cyber
Security (ACCS) used the \textit{IXIA PerfectStorm} tool to create a combination
of normal and abnormal network traffic. The abnormal traffic generated from the
IXIA PerfectStorm tool is broken down into nine attack types, which we used as our
source classes for generating our adversarial examples\footnotemark. After the
traffic is generated, the authors leveraged \textit{Argus} and \textit{Bro-IDS}
tools to construct reliable features. 

\footnotetext{While the authors intended this dataset to be used for benchmarking
\textit{anomaly} detection algorithms, we used it as a \textit{signature-based}
dataset by using the last feature, ``attack class,'' as the label. We achieved
high classification accuracies, and thus argue that our modification has no impact
on the significance of our results found in the evaluation of our methodology.}

The UNSW-NB15 contains 10 classes, with 9 attack classes and 1 benign class. It
contains 175,341 samples for training and 83,332 samples for testing. The dataset
contains 48 features, separated into four high-level categories of features:
flow-based, basic connection, content, and time-based. We defer to the authors for
a comprehensive description of the dataset~\cite{ms15unsw} and its similarity with
the NSL-KDD~\cite{mt16evaluation,ms17hybrid}.

\vspace{3pt}\noindent\textbf{MNIST.} The Modified National Institute of Standards
and Technology database contains hanwritten digits. We chose to use MNIST due to
its simplicity, to demonstrate cross-domain applicability of our approach, and
to have a direct comparison with other adversarial machine learning approaches.

The MNIST database contains 10 classes, with numerical digits from 0 through 9.
It contains 60,000 samples for training and 10,000 samples for testing. Unlike the
network intrusion detection datasets, no preprocessing was required to integrate
this dataset into our experimental setup. We were able to use the dataset directly
as-is. We defer to the authors for the intricate details surrounding the MNIST
dataset~\cite{lc10mnist}.

\looseness=-1
\vspace{3pt}\noindent\textbf{GTSRB.} The German Traffic Sign Recognition Benchmark
is a dataset of common traffic signs found throughout Germany. We chose to use the
GTSRB to for its increased complexity over MNIST and as a second example of the
cross-domain applicability of our methodology.

\looseness=-1
The GTSRB contains 42 classes. After preprocessing, our experiments contained
21,792 samples for training and 6,893 samples for testing. Throughout the dataset,
there are identical images of varying sizes. We first cropped the region of
interest (which contains the traffic sign) and downsampled to a final size of
$30\times30$. For more details concerning the GTSRB, we defer to the
authors~\cite{sk12mvc}.

\subsection{Experiment Overview} In this subsection, we describe our experiments
in detail, using the NSL-KDD as an application of our approach.

\looseness=-1 Through our evaluation on network intrusion detection, we note that
our experiments emulate a realistic adversary by taking different attacks and
masking them as benign traffic.

\vspace{3pt}\noindent\textbf{Data Curation.} We perform a stratified shuffle-split
on the training set into five parts. Splitting our training set into five parts
lays the foundation for measuring transferability: each partition (which we refer
to as \textit{A, B, C, D, E}) is representative of a uniquely trained model (which
we refer to as \textit{$M_A$, $M_B$, $M_C$, $M_D$, $M_E$} respectively), mirroring
the setup described in ~\cite{pp16transferability}. This setup allows us to
measure \textit{intra-technique} transferability rates: the rate at which
adversarial examples crafted from one model are also misclassified by another
model with the same learning technique. Furthermore, we can also measure the
converse, \textit{inter-technique} transferability: the same misclassification
rate except with a model of an entirely different learning technique\footnotemark.
Note, that \textit{transferability} is not the same as a black-box attack; we do
not use the output of an oracle to label data for a surrogate model, as shown
in~\cite{pp16transferability,iy18black}. While transferability \textit{enables}
black-box attacks, they are not the same, as transferability has been observed
prior to the inception of black-box attacks~\cite{sg13intriguing,gf15explain}.
 
\footnotetext{\looseness=-1 For our inter-technique evaluation, we consider four popular
learning techniques: Logistic Regression (LR), Decision Trees (DT), Support Vector
Machine (SVM), and K-Nearest Neighbors (KNN). Each one of these learners
represent different learning paradigms (and are popular in commercial and academic
contexts), and are thus appropriate candidates for evaluating inter-technique
transferability. Hyperparameters and other details can be found in the
\hyperref[sec:appendix]{Appendix}.}

\looseness=-1 Next, we perform a stratified shuffle-split on the test set. This
creates two sets of isolated inputs: we use the \ajsma{} on one set and the \hsg{}
on the other.  This is to ensure that the sketches are only applied on inputs that
had no influence in the creation of the sketch (recall that we analyze the
adversarial examples created by the \ajsma{} to build sketches). Furthermore, the
stratified shuffle-split of the test set is to ensure that the \ajsma{} crafts
adversarial examples from inputs spanning all classes. This enables us to create
effective adversarial sketches, as all class-specific information is distilled
into the perturbation histogram.

\looseness=-1 Using the NSL-KDD as an example for the first stage, we built a
Multi-layer Perceptron with 4 layers: an input layer of 123 units, fully-connected
to 64 units, fully-connected to 32 units, and finally an output layer with 5
units\footnotemark. The output layer conveys our 5 classes: Normal (Benign),
Probe, Denial of Service (DoS), User to Root (U2R), and Remote to Local (R2L). We
used rectified linear units (ReLU) as our chosen activation function for our
hidden layers and softmax at the output layer. Our models are trained via the Adam
optimizer~\cite{km14adam} with a batch size of 200 and a learning rate of 0.01 for
5 epochs. With our five splits, each model, $M_A$ through $M_E$, is trained with
$\mathtt{\sim}$25,194 inputs. With these hyperparameters, network architecture,
and training set size, we were able to achieve an average $77\%\pm 1.0\%$ accuracy
on the test set, which is consistent with the literature~\cite{tl09detailed,
ig15performance,ih13comparison,ms17hybrid}.

\footnotetext{\looseness=-1 We note that the number of layers and units was influenced by
research that suggests an optimal upper bound for the number hidden neurons for
feed-forward networks~\cite{hg98upper}. The remainder of our hyperparameter
selection follows no formal process.}

\looseness=-1
\vspace{3pt}\noindent\textbf{Constraint Generation.} With the heuristic described
in \S\ref{sec:methodology}, we learn constraints in the NSL-KDD with
\textsc{protocol} as the primary feature. The intuition behind this selection is
straightforward: a majority of the features in the NSL-KDD describe metadata
surrounding these protocols, e.g., flag information, services, and content-related
features like \textsc{FTP} commands. The distribution of the extracted constraints
for the NSL-KDD can be found in Table \ref{tab:nslconst}. We observe that the TCP
protocol offers the highest degree of maneuverability by a wide margin (and to no
surprise as it constitutes the majority of traffic flows in the dataset). This is
unlike UDP and ICMP, who are significantly more constrained. Table
\ref{tab:nsl-constraints} in the \hyperref[sec:appendix]{Appendix} shows all of
the constraints, sorted by feature type.

\begin{table}[!t]
  \resizebox{\columnwidth}{!}{ \centering 
  \begin{tabular}{lccccr}
    \toprule

    {} & \multicolumn{4}{c}{Feature Type} & {}\\
    \cmidrule{2-5}
    {Protocol} & {Basic} & {Content} & {Timing-based} & {Host-based} & {Total}\\
    \midrule
    {\textbf{TCP}} & {81} & {12} & {9} & {10} & {112}\\
    {\textbf{UDP}} & {12} & {0} & {7} & {8} & {27}\\
    {\textbf{ICMP}} & {14} & {0} & {7} & {8} & {29}\\

    \bottomrule
  \end{tabular}}

  \caption{NSL-KDD constraint distribution, categorized by feature type - Unlike
  TCP, UDP and ICMP have limited degrees of freedom.} \label{tab:nslconst}

\end{table}

\begin{figure*}
  \includegraphics[width=\textwidth]{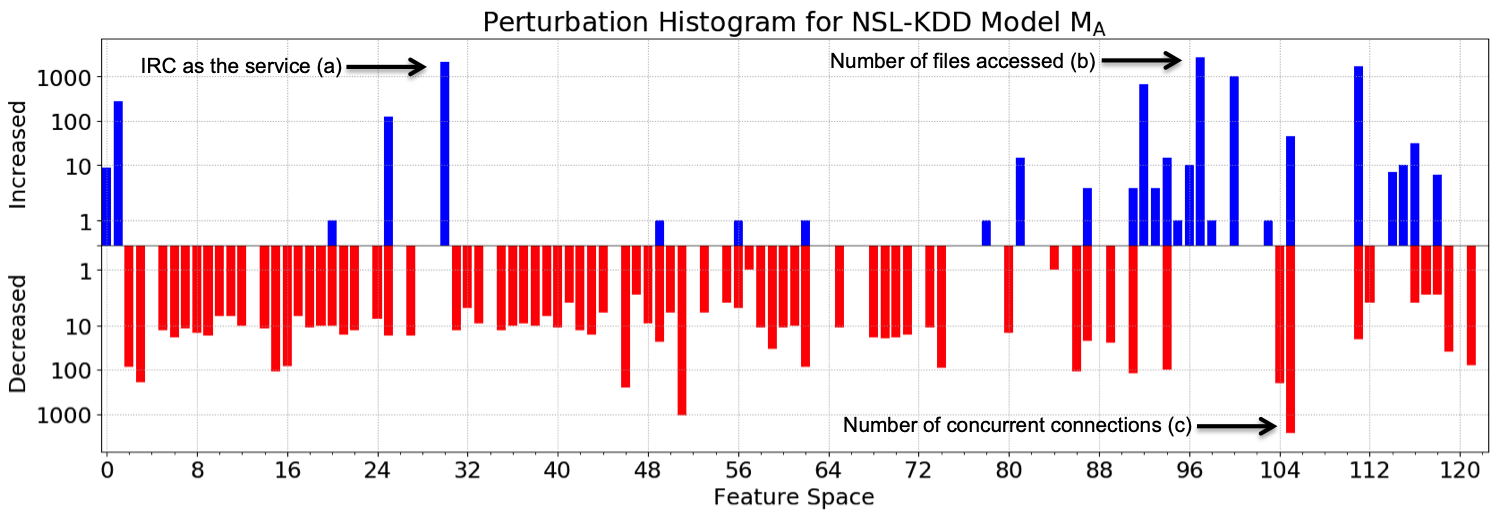}

  \caption{NSL-KDD model $M_B$ Perturbation Histogram produced by adversarial
  examples from the \ajsma{} in log scale - Certain features are
  \textit{consistently} increased (a) \& (b) and decreased (c), indifferent of
  the source class.} \label{fig:nslhisto}

\end{figure*}

\vspace{3pt}\noindent\textbf{Adversarial Example Generation.} To investigate the
first question in our evaluation, we craft adversarial examples with the \ajsma{}.
With the constraints integrated into the crafting process, we iterate over the
first half of the test set for each model, $M_A$ through $M_E$, and craft
adversarial examples (the second half of the test set is used to craft adversarial
examples via the sketches produced by the \hsg{}). Table \ref{tab:nslcraft}
describes the output of this stage from an example NSL-KDD run with ``Benign'' (0)
as the target class. Figure \ref{fig:nslhisto} demonstrates the perturbation
histogram computed from the same run.

\begin{table}[!t]
  \resizebox{\columnwidth}{!}{ \centering 
  \begin{tabular}{lc}
    \toprule

    \multicolumn{2}{c}{\ajsma{} Experiment Results - Target 0 ``Benign''}\\
    \midrule
    {Testing Inputs} & {11,272}\\
    {Labeled as Target Class} & {4,856}\\
    {Misclassified as Target Class} & {2,532}\\
    {Number of Inputs Attacked} & {3,884}\\
    {Average Distortion} & {3.39\% $\mathtt{\sim}$4.18 features}\\
    {Class Success Rates} & 0:NaN, 1:100\%, 2:99\%\\
		{} & {3.99\%, 4:100\%}\\

    \bottomrule
  \end{tabular}}

  \caption{Output from crafting adversarial examples with the \ajsma{} for NSL-KDD
  model, $M_B$ - Even domains with constraints are vulnerable to adversarial
  examples.} \label{tab:nslcraft}

\end{table}

\begin{figure}[t]

  \includegraphics[width=0.49\columnwidth]{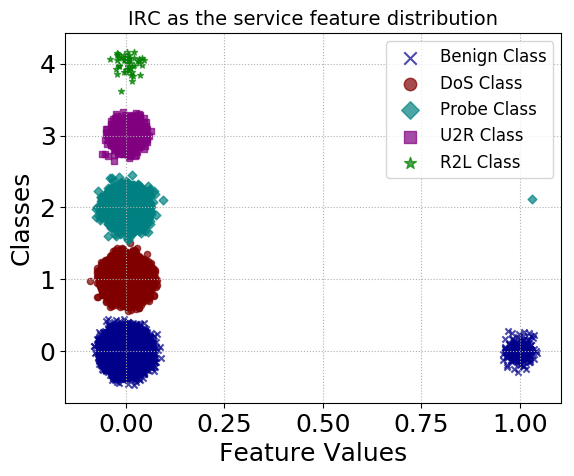}
  \includegraphics[width=0.49\columnwidth]{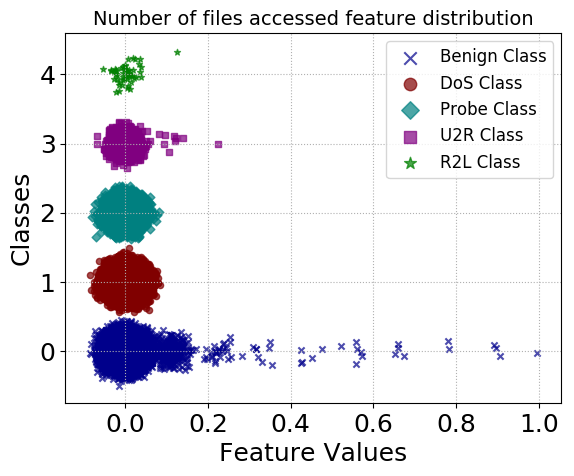}

  \caption{NSL-KDD class distribution for \textsc{service=IRC} (a) and
  \textsc{num\_access\_files} (b) - These two features are often perturbed due to
  their bias towards the ``Benign'' class, shown as class 0 on the Y-axis.} \label{fig:irc-naf}

\end{figure}

\looseness=-1
We note that there are particular features that were consistently perturbed in nearly
all adversarial examples, namely setting the service as \textsc{IRC} and
increasing \textsc{num\_access\_files}\footnotemark. To understand why, we used
WEKA~\cite{hl09weka} to analyze the distribution of these features, which revealed
a trivial explanation: inputs that use \textsc{IRC} as the service and have high
values for \textsc{num\_access\_files} are heavily skewed towards our target
class, ``Benign''. Figure \ref{fig:irc-naf} shows these distributions (Note that
the ``Benign'' target class is bottom class (0) on the Y-axis).

\footnotetext{While it may seem this feature is not derived from network data,
both the NSL-KDD and UNSW-NB15 have a ``high-level content`` category of features,
which are derrived from payload information.}

\looseness=-1 \vspace{3pt}\noindent\textbf{Adversarial Sketch Generation.} To
investigate the second question in our evaluation, we use the perturbation
histogram (computed from adversarial examples crafted by the \ajsma{}) to create
adversarial sketches. As described in \S\ref{sec:methodology}, the \hsg{} creates
adversarial sketches by selecting the top $n$ most perturbed features from the
perturbation histogram. Finally, we craft adversarial examples by applying the
sketch to the second half of the test set, for each model, $M_A$ through $M_E$.

\looseness=-1 \subsection{Measuring Success} With the adversarial examples crafted
via the \ajsma{} and \hsg{}, we measure the effectiveness of the two algorithms
through white-box attacks and through transferability. We define the success rate
of white-box attacks as the number of adversarial examples misclassified as the
target class over the total number of attempted inputs, formally:
\[ SR_{wb} = \frac{|\{\vect{x} \in \mathbb{X}: f(\vect{x}) = t\}|}{|\mathbb{X}|}\]
\looseness=-1
where $\mathbb{X}$ represents the set of attempted inputs, $f$ is a model, and $t$
is the target class. Furthermore, we define the transferability success rate to be
the number of adversarial examples misclassified as the target class by the target
model over the number of successful adversarial examples crafted from the source
model, formally:
\[ SR_{transfer} = \frac{|\{\vect{x} \in \mathbb{X}: f'(\vect{x}) = t\}|}
{|\{\vect{x} \in \mathbb{X}: f(\vect{x}) = t\}|} \]
where $\mathbb{X}$ again represents the set of attempted inputs from the source
model $f$, and $f'$ represents the target model.

\vspace{3pt}\noindent\textbf{AJSMA Results.} In Table \ref{tab:results} (a) on the
left, we show the NSL-KDD \ajsma{} success rates for white-box attacks and
transferability rates. The labels on the left represent the source model used to
generate adversarial examples and the labels on top represent the target models
that were attacked. For the intra-technique case, the white-box results can be
read along the diagonal (as the source and target are the same model). For both
intra- and inter-technique cases, the transferability rates are represented in
all other cells. The \ajsma{} was broadly successful in creating targeted
adversarial examples while introducing relative amounts of distortion comparable
to image-based experiments, even in the presence of constraints. These results
suggest that the constraints for our evaluated domain do not offer any robustness
against adversarial examples.

\looseness=-1
Additionally, it is interesting that the adversarial examples produced by the
\ajsma{} had notable transferability rates in intra-technique case (an average of
73\% across our network intrusion detection experiments). This would suggest that
transferability is stronger in lower dimensional spaces.

\begin{figure}[!t]
  \includegraphics[width=\columnwidth]{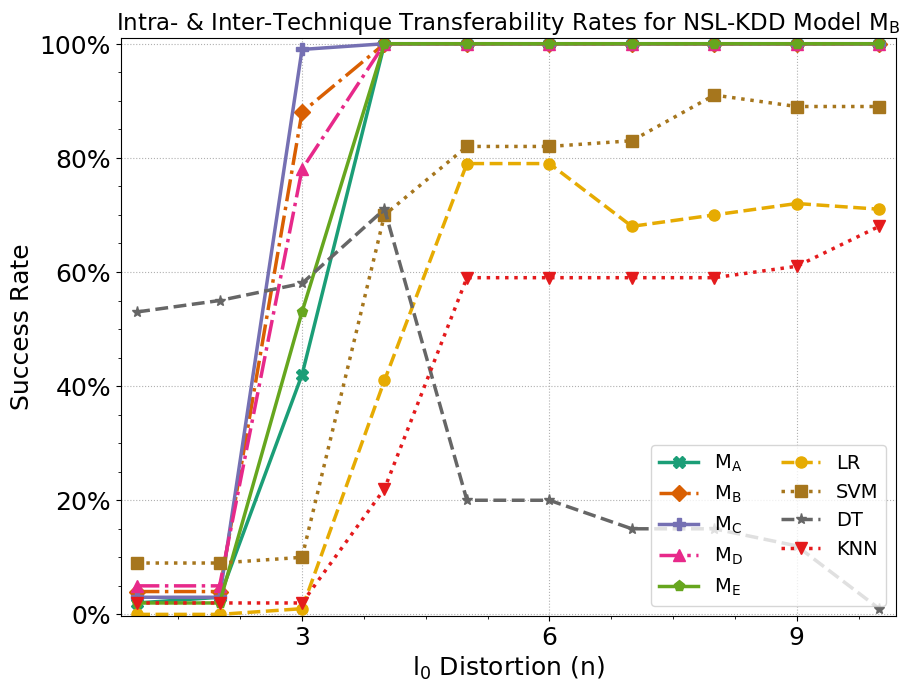}

   \caption{\hsg{} intra- and inter- transferability rates for NSL-KDD model
   $M_B$ for target ``Benign'' as a function of $l_0$ distortion $n$ - Values
   between 6-9 for $n$ have greater than 70\% misclassification for most learning
   techniques.} \label{fig:nslbbplot}

\end{figure}

\vspace{3pt}\noindent\textbf{HSG Results.} In Figure \ref{fig:nslbbplot}, we show
NSL-KDD model $M_B$ success rates for intra- and inter-technique transferability
for varying values of $n$ between 1 and 11. There are broad regions for values of
$n$ which have high success rates, reaching 100\% in white-box settings and
greater than 80\% transferability rates for the majority of learning techniques.
Indeed, it would appear that the most often perturbed features (by the \ajsma{})
are appropriate candidates for building an adversarial sketch. Furthermore, these
results confirm the existence of universal adversarial perturbations that obey
domain constraints.

\looseness=-1
Finally, we are surprised at the fragility of the models trained on network
intrusion detection data; sometimes perturbing only \textit{three} features was
needed to misclassify greater than 80\% of inputs in the test set. We believe that
this fragility is partly a function of the skewed distribution certain features
can have for specific classes, such as the ones shown in Figure \ref{fig:irc-naf}.
This insight is unlike the adversarial examples crafted in image domains, where
high dimensionality and a more balanced class distribution for features appear to
mitigate this fragility. 

Table \ref{tab:results} shows the success rates for both of our algorithms across
all four datasets. The values of $n$ listed equal $\mathtt{\sim}4\%$ $l_0$
distortion for the \hsg{}. Again, the labels on the left represent the source
model used to generate adversarial examples, while the labels on top represent the
target models that were attacked. For the intra-technique case, the white-box
results can be read along the diagonal (as the source and target models are the
same). The transferability rates, for both intra- and inter-technique cases, are
represented in all other cells. 

\looseness=-1
We observe high success rates for both the \ajsma{} and the \hsg{} in both
white-box attacks and transferability for several of the datasets. While the
\ajsma{} struggled significantly to produce adversarial examples that transferred
as the dimensionality increased, the \hsg{} transferability rates were more
resilient (but still affected) to the increased dimensionality\footnotemark. This
would suggest that the distance between decision boundaries increases as model
complexity increases, thus mitigating transferability for our algorithms.

\footnotetext{We hypothesize that this is partly due to how the \ajsma{} is
designed. Conceptually, the \ajsma{} creates adversarial examples that
\textit{just} cross over the decision boundaries. While the decision boundaries
among models of lower dimensionality may be similar (and thus, an adversarial
example can cross over the decision boundaries of multiple models), this appears
to be untrue for models of higher dimensionality, where there can be multiple
unique ways to separate the data. We could test this by regularizing high
dimensional models. Furthermore, we could also alter the design of the \ajsma{} to
continue adding additional perturbations so that it returns an adversarial example
that is more confidently misclassified.}

\newcommand{\ajsmakdd}{%
\resizebox{0.5\columnwidth}{!}{ \centering 
\begin{tabular}{c c}

  \multicolumn{1}{c}{\normalsize \textsc{Adaptive JSMA}} \\
  \cmidrule(l{11em}r{11.05em}){1-1}

  \begin{tabular}{c|c|c|c|c|c|}
    \mca{} & \mca{$M_A$}  & \mca{$M_B$} & \mca{$M_C$} & \mca{$M_D$} & \mca{$M_E$} \\ \hhline{*1{~}*{5}{|-}}
    \mcb{$M_A$} & \cc{100} & \cc{69} & \cc{51}  & \cc{51} & \cc{61}  \\ \hhline{*1{~}*{5}{|-}}
    \mcb{$M_B$} & \cc{73}  & \cc{99} & \cc{72}  & \cc{78} & \cc{67}  \\ \hhline{*1{~}*{5}{|-}}
    \mcb{$M_C$} & \cc{63}  & \cc{69} & \cc{100} & \cc{66} & \cc{69}  \\ \hhline{*1{~}*{5}{|-}}
    \mcb{$M_D$} & \cc{54}  & \cc{93} & \cc{70}  & \cc{99} & \cc{64}  \\ \hhline{*1{~}*{5}{|-}}
    \mcb{$M_E$} & \cc{83}  & \cc{79} & \cc{71}  & \cc{66} & \cc{100} \\ \hhline{*1{~}*{5}{|-}}
  \end{tabular}
  
  \begin{tabular}{|c|c|c|c|}
    \mca{$LR$} & \mca{$SVM$} & \mca{$DT$} & \mca{$KNN$} \\ \hhline{*{4}{|-}}
    \cc{34}  & \cc{40}  &  \cc{34}  & \cc{20} \\ \hhline{*{4}{|-}}
    \cc{49}  & \cc{50}  &  \cc{39}  & \cc{44} \\ \hhline{*{4}{|-}}
    \cc{30}  & \cc{34}  &  \cc{27}  & \cc{24} \\ \hhline{*{4}{|-}}
    \cc{24}  & \cc{38}  &  \cc{25}  & \cc{22} \\ \hhline{*{4}{|-}}
    \cc{42}  & \cc{55}  &  \cc{41}  & \cc{17} \\ \hhline{*{4}{|-}}
  \end{tabular}

\end{tabular}}}

\newcommand{\ajsmaunsw}{%
\resizebox{0.5\columnwidth}{!}{ \centering 
\begin{tabular}{c c}
  \begin{tabular}{c|c|c|c|c|c|}
    \mca{} & \mca{$M_A$}  & \mca{$M_B$} & \mca{$M_C$} & \mca{$M_D$} & \mca{$M_E$} \\ \hhline{*1{~}*{5}{|-}}
    \mcb{$M_A$} & \cc{100} & \cc{97}  & \cc{92}  & \cc{96}  & \cc{96}  \\ \hhline{*1{~}*{5}{|-}}
    \mcb{$M_B$} & \cc{50}  & \cc{100} & \cc{72}  & \cc{94}  & \cc{71}  \\ \hhline{*1{~}*{5}{|-}}
    \mcb{$M_C$} & \cc{73}  & \cc{81}  & \cc{100} & \cc{93}  & \cc{87}  \\ \hhline{*1{~}*{5}{|-}}
    \mcb{$M_D$} & \cc{69}  & \cc{64}  & \cc{55}  & \cc{100} & \cc{59}  \\ \hhline{*1{~}*{5}{|-}}
    \mcb{$M_E$} & \cc{66}  & \cc{80}  & \cc{90}  & \cc{96}  & \cc{100} \\ \hhline{*1{~}*{5}{|-}}
  \end{tabular}

  \begin{tabular}{|c|c|c|c|}
    \mca{$LR$} & \mca{$SVM$} & \mca{$DT$} & \mca{$KNN$} \\ \hhline{*{4}{|-}}
    \cc{72} & \cc{81} & \cc{29} & \cc{53} \\ \hhline{*{4}{|-}}
    \cc{62} & \cc{62} & \cc{12} & \cc{26} \\ \hhline{*{4}{|-}}
    \cc{71} & \cc{76} & \cc{19} & \cc{49} \\ \hhline{*{4}{|-}}
    \cc{53} & \cc{48} & \cc{8}  & \cc{25} \\ \hhline{*{4}{|-}}
    \cc{66} & \cc{69} & \cc{15} & \cc{38} \\ \hhline{*{4}{|-}}
  \end{tabular}
\end{tabular}}}

\newcommand{\ajsmamnist}{%
\resizebox{0.5\columnwidth}{!}{ \centering 
\begin{tabular}{c c}
  \begin{tabular}{c|c|c|c|c|c|}
    \mca{} & \mca{$M_A$}   & \mca{$M_B$} & \mca{$M_C$} & \mca{$M_D$} & \mca{$M_E$} \\ \hhline{*1{~}*{5}{|-}}
    \mcb{$M_A$} & \cc{100} & \cc{39}  & \cc{30} & \cc{19} & \cc{11}  \\ \hhline{*1{~}*{5}{|-}}
    \mcb{$M_B$} & \cc{6}   & \cc{100} & \cc{14} & \cc{6}  & \cc{2}  \\ \hhline{*1{~}*{5}{|-}}
    \mcb{$M_C$} & \cc{12}  & \cc{32} & \cc{99}  & \cc{9}  & \cc{8}  \\ \hhline{*1{~}*{5}{|-}}
    \mcb{$M_D$} & \cc{17}  & \cc{44} & \cc{26}  & \cc{100} & \cc{21} \\ \hhline{*1{~}*{5}{|-}}
    \mcb{$M_E$} & \cc{20}  & \cc{53} & \cc{33}  & \cc{20} & \cc{99} \\ \hhline{*1{~}*{5}{|-}}
  \end{tabular}

  \begin{tabular}{|c|c|c|c|}
    \mca{$LR$} & \mca{$SVM$} & \mca{$DT$} & \mca{$KNN$} \\ \hhline{*{4}{|-}}
    \cc{21} & \cc{24} & \cc{18} & \cc{5} \\ \hhline{*{4}{|-}}
    \cc{19} & \cc{18} & \cc{14} & \cc{2} \\ \hhline{*{4}{|-}}
    \cc{23} & \cc{22} & \cc{15} & \cc{1} \\ \hhline{*{4}{|-}}
    \cc{21} & \cc{21} & \cc{19} & \cc{5} \\ \hhline{*{4}{|-}}
    \cc{24} & \cc{27} & \cc{18} & \cc{6} \\ \hhline{*{4}{|-}}
  \end{tabular}
\end{tabular}}}

\newcommand{\ajsmagtsrb}{%
\resizebox{0.5\columnwidth}{!}{ \centering 
\begin{tabular}{c c}
  \begin{tabular}{c|c|c|c|c|c|}
    \mca{} & \mca{$M_A$}  & \mca{$M_B$} & \mca{$M_C$} & \mca{$M_D$} & \mca{$M_E$} \\ \hhline{*1{~}*{5}{|-}}
    \mcb{$M_A$} & \cc{97}  & \cc{2}   & \cc{1}   & \cc{3}  & \cc{2}   \\ \hhline{*1{~}*{5}{|-}}
    \mcb{$M_B$} & \cc{1}   & \cc{95}  & \cc{0.5} & \cc{1}  & \cc{1}   \\ \hhline{*1{~}*{5}{|-}}
    \mcb{$M_C$} & \cc{2}   & \cc{2}   & \cc{98}  & \cc{2}  & \cc{0.8} \\ \hhline{*1{~}*{5}{|-}}
    \mcb{$M_D$} & \cc{0.5} & \cc{1}   & \cc{3}   & \cc{99} & \cc{1}   \\ \hhline{*1{~}*{5}{|-}}
    \mcb{$M_E$} & \cc{0.7} & \cc{2}   & \cc{1}   & \cc{2}  & \cc{94}  \\ \hhline{*1{~}*{5}{|-}}
  \end{tabular}

  \begin{tabular}{|c|c|c|c|}
    \mca{$LR$} & \mca{$SVM$} & \mca{$DT$} & \mca{$KNN$} \\ \hhline{*{4}{|-}}
    \cc{0.2} & \cc{1}   & \cc{3} & \cc{0} \\ \hhline{*{4}{|-}}
    \cc{0.3} & \cc{2}   & \cc{2} & \cc{0} \\ \hhline{*{4}{|-}}
    \cc{0.1} & \cc{2}   & \cc{1} & \cc{0} \\ \hhline{*{4}{|-}}
    \cc{0.2} & \cc{0.8} & \cc{3} & \cc{0} \\ \hhline{*{4}{|-}}
    \cc{0.1} & \cc{0.5} & \cc{2} & \cc{0} \\ \hhline{*{4}{|-}}
  \end{tabular}
\end{tabular}}}

\newcommand{\topnkdd}{%
\resizebox{0.5\columnwidth}{!}{ \centering 
\begin{tabular}{c c}

  \multicolumn{1}{c}{\large \textsc{Histogram Sketch Generation}} \\
  \cmidrule(l{8.5em}r{8.5em}){1-1}

  \begin{tabular}{c|c|c|c|c|c|}
    \mcr{$n=6$} & \mca{$M_A$}  & \mca{$M_B$} & \mca{$M_C$} & \mca{$M_D$} & \mca{$M_E$} \\ \hhline{*1{~}*{5}{|-}}
    \mcb{$M_A$} & \cc{100} & \cc{100} & \cc{100} & \cc{100} & \cc{100} \\ \hhline{*1{~}*{5}{|-}}
    \mcb{$M_B$} & \cc{100} & \cc{100} & \cc{100} & \cc{100} & \cc{100} \\ \hhline{*1{~}*{5}{|-}}
    \mcb{$M_C$} & \cc{100} & \cc{100} & \cc{100} & \cc{100} & \cc{100} \\ \hhline{*1{~}*{5}{|-}}
    \mcb{$M_D$} & \cc{100} & \cc{100} & \cc{100} & \cc{100} & \cc{100} \\ \hhline{*1{~}*{5}{|-}}
    \mcb{$M_E$} & \cc{100} & \cc{100} & \cc{100} & \cc{100} & \cc{100} \\ \hhline{*1{~}*{5}{|-}}
  \end{tabular}

  \begin{tabular}{|c|c|c|c|}
    \mca{$LR$} & \mca{$SVM$} & \mca{$DT$} & \mca{$KNN$} \\ \hhline{*{4}{|-}}
    \cc{80}  & \cc{85}  &  \cc{16}  & \cc{19} \\ \hhline{*{4}{|-}}
    \cc{79}  & \cc{82}  &  \cc{20}  & \cc{59} \\ \hhline{*{4}{|-}}
    \cc{65}  & \cc{83}  &  \cc{50}  & \cc{34} \\ \hhline{*{4}{|-}}
    \cc{81}  & \cc{87}  &  \cc{22}  & \cc{59} \\ \hhline{*{4}{|-}}
    \cc{98}  & \cc{93}  &  \cc{27}  & \cc{19} \\ \hhline{*{4}{|-}}
  \end{tabular}

\end{tabular}}}

\newcommand{\topnunsw}{%
\resizebox{0.5\columnwidth}{!}{ \centering 
\begin{tabular}{c c}
  \begin{tabular}{c|c|c|c|c|c|}
    \mcr{$n=9$} & \mca{$M_A$}  & \mca{$M_B$} & \mca{$M_C$} & \mca{$M_D$} & \mca{$M_E$} \\ \hhline{*1{~}*{5}{|-}}
    \mcb{$M_A$} & \cc{88} & \cc{99}  & \cc{95}  & \cc{96}  & \cc{96} \\ \hhline{*1{~}*{5}{|-}}
    \mcb{$M_B$} & \cc{99} & \cc{100} & \cc{99}  & \cc{100} & \cc{99} \\ \hhline{*1{~}*{5}{|-}}
    \mcb{$M_C$} & \cc{93} & \cc{97}  & \cc{100} & \cc{77}  & \cc{95} \\ \hhline{*1{~}*{5}{|-}}
    \mcb{$M_D$} & \cc{80} & \cc{99}  & \cc{100} & \cc{100} & \cc{99} \\ \hhline{*1{~}*{5}{|-}}
    \mcb{$M_E$} & \cc{98} & \cc{100} & \cc{100} & \cc{94}  & \cc{92} \\ \hhline{*1{~}*{5}{|-}}
  \end{tabular}

  \begin{tabular}{|c|c|c|c|}
    \mca{$LR$} & \mca{$SVM$} & \mca{$DT$} & \mca{$KNN$} \\ \hhline{*{4}{|-}}
    \cc{99}  & \cc{96}  & \cc{29} & \cc{37} \\ \hhline{*{4}{|-}}
    \cc{99}  & \cc{100} & \cc{20} & \cc{62} \\ \hhline{*{4}{|-}}
    \cc{73}  & \cc{100} & \cc{27} & \cc{41} \\ \hhline{*{4}{|-}}
    \cc{74}  & \cc{99}  & \cc{13} & \cc{30} \\ \hhline{*{4}{|-}}
    \cc{85}  & \cc{100} & \cc{28} & \cc{39} \\ \hhline{*{4}{|-}}
  \end{tabular}
\end{tabular}}}

\newcommand{\topnmnist}{%
\resizebox{0.5\columnwidth}{!}{ \centering 
\begin{tabular}{c c}
  \begin{tabular}{c|c|c|c|c|c|}
    \mcr{$n=41$} & \mca{$M_A$}  & \mca{$M_B$} & \mca{$M_C$} & \mca{$M_D$} & \mca{$M_E$} \\ \hhline{*1{~}*{5}{|-}}
    \mcb{$M_A$} & \cc{24} & \cc{35} & \cc{18} & \cc{14} & \cc{9}  \\ \hhline{*1{~}*{5}{|-}}
    \mcb{$M_B$} & \cc{17} & \cc{42} & \cc{14} & \cc{15} & \cc{54} \\ \hhline{*1{~}*{5}{|-}}
    \mcb{$M_C$} & \cc{22} & \cc{31} & \cc{41} & \cc{9}  & \cc{14} \\ \hhline{*1{~}*{5}{|-}}
    \mcb{$M_D$} & \cc{13} & \cc{30} & \cc{14} & \cc{36} & \cc{13} \\ \hhline{*1{~}*{5}{|-}}
    \mcb{$M_E$} & \cc{15} & \cc{29} & \cc{11} & \cc{12} & \cc{31} \\ \hhline{*1{~}*{5}{|-}}
  \end{tabular}

  \begin{tabular}{|c|c|c|c|}
    \mca{$LR$} & \mca{$SVM$} & \mca{$DT$} & \mca{$KNN$} \\ \hhline{*{4}{|-}}
    \cc{45} &\cc{48} & \cc{14} & \cc{7} \\ \hhline{*{4}{|-}}
    \cc{62} &\cc{61} & \cc{13} & \cc{5} \\ \hhline{*{4}{|-}}
    \cc{60} &\cc{55} & \cc{32} & \cc{5} \\ \hhline{*{4}{|-}}
    \cc{46} &\cc{46} & \cc{33} & \cc{6} \\ \hhline{*{4}{|-}}
    \cc{44} &\cc{58} & \cc{21} & \cc{4} \\ \hhline{*{4}{|-}}
  \end{tabular}
\end{tabular}}}

\newcommand{\topngtsrb}{%
\resizebox{0.5\columnwidth}{!}{ \centering 
\begin{tabular}{c c}
  \begin{tabular}{c|c|c|c|c|c|}
    \mcr{$n=104$} & \mca{$M_A$}  & \mca{$M_B$} & \mca{$M_C$} & \mca{$M_D$} & \mca{$M_E$} \\ \hhline{*1{~}*{5}{|-}}
    \mcb{$M_A$} & \cc{34} & \cc{6} & \cc{27} & \cc{6} & \cc{7} \\ \hhline{*1{~}*{5}{|-}}
    \mcb{$M_B$} & \cc{9} & \cc{27} & \cc{14} & \cc{2} & \cc{8} \\ \hhline{*1{~}*{5}{|-}}
    \mcb{$M_C$} & \cc{17} & \cc{4} & \cc{44} & \cc{2} & \cc{7} \\ \hhline{*1{~}*{5}{|-}}
    \mcb{$M_D$} & \cc{29} & \cc{12} & \cc{16} & \cc{29} & \cc{12} \\ \hhline{*1{~}*{5}{|-}}
    \mcb{$M_E$} & \cc{14}  & \cc{8} & \cc{15} & \cc{3} & \cc{35} \\ \hhline{*1{~}*{5}{|-}}
  \end{tabular}

  \begin{tabular}{|c|c|c|c|}
    \mca{$LR$} & \mca{$SVM$} & \mca{$DT$} & \mca{$KNN$} \\ \hhline{*{4}{|-}}
    \cc{0.2}   & \cc{2} & \cc{12} & \cc{0} \\ \hhline{*{4}{|-}}
    \cc{0.5} & \cc{1} & \cc{6}  & \cc{0} \\ \hhline{*{4}{|-}}
    \cc{0.6}   & \cc{1} & \cc{5}  & \cc{0} \\ \hhline{*{4}{|-}}
    \cc{1}   & \cc{1} & \cc{5}  & \cc{0} \\ \hhline{*{4}{|-}}
    \cc{0.1}   & \cc{1} & \cc{4}  & \cc{0} \\ \hhline{*{4}{|-}}
  \end{tabular}
\end{tabular}}}

\begin{table*}[ht]
  \centering 
  \begin{subtable}[h]{\textwidth}
    \ajsmakdd
    \topnkdd
    \caption{NSL-KDD}
  \end{subtable}
  \begin{subtable}[h]{\textwidth}
    \ajsmaunsw
   \topnunsw
   \caption{UNSW-NB15}
  \end{subtable}
  \begin{subtable}[h]{\textwidth}
    \ajsmamnist
   \topnmnist
   \caption{MNIST}
  \end{subtable}
  \begin{subtable}[h]{\textwidth}
   \ajsmagtsrb
   \topngtsrb
   \caption{GTSRB}
 \end{subtable}

  \caption{Results for \ajsma{} (left) and \hsg{} (right) for all of our
  experiments for target class ``Benign'' - Values of $n$ for our sketches
  represent  $\mathtt{\sim}4\%$ $l_0$ distortion.}
  
  \label{tab:results}
\end{table*}

From our investigation, we highlight some key takeaways: 

\begin{enumerate} 

  \item Constraints, as they stand, are not problematic for crafting adversarial
    examples in the domains we studied. The \ajsma{} reached $100\%$ success rates
    for most attempted inputs, with distortion rates comparable to image-based
    experiments, while in the presence of constraints.
  
  \item Universal adversarial perturbations that obey domain constraints exist.
    The \hsg{} produced Adversarial Sketches that reached 100\% success rates in
    constrained domains for a values of $n$ that represent $\mathtt{\sim}4\%$
    $l_0$ distortion. Furthermore, they produced adversarial examples that had
    higher transferability rates than the \ajsma{}.

  \item Network intrusion detection data is highly fragile: a small dimensionality
    and biased distributions enable attack algorithms to alter very few features
    to successfully craft targeted adversarial examples.
  
  \item Worst-case scenarios (i.e., white-box attacks) are highly vulnerable, not
    surprisingly. With direct access to model parameters, an adversary can have a
    sophisticated level of control over the output of a model.

  \item Network intrusion detection data appears to be highly vulnerable to
    transferability attacks. Even in the presence of disjoint training sets and
    different learning techniques, both attacks produced adversarial examples with
    surprising levels (an average of 73\% for the \ajsma{} and 97\% for the
    \hsg{}, for models of the same learning technique) of transferability.

\end{enumerate}

\section{Uncontrollable Features} \label{sec:uncontrollable} Throughout this
paper, we have discussed some of the differences between adversarial machine
learning in constrained domains versus unconstrained domains. One of the most
fundamental questions within this community is: \textit{what precisely is an
``adversarial example?''} There are varying definitions with different objectives
used throughout AML research. In the image space, it has been generally agreed
that if an attack algorithm produces perturbations that are undetectable by a
human observer, then it is an adversarial example. However, it is not clear how to
translate this objective in other domains.

\looseness=-1 Research outside of image space (usually) provides their own
definitions: perturbed malware must maintain its properties of
malware~\cite{gs17adversarial,kj18adversarial}, perturbed audio must be nearly
inaudible~\cite{cl18audio}, perturbed text must preserve its
semantics~\cite{eh18hotflip,ji17adversarial}, among other definitions. For our
work in network intrusion detection, we follow an intuitive definition: perturbed
network flows must maintain their attack behavior. For example, a DoS attack must
still be a DoS attack post-perturbation.

However, validating attack behaviors is a nontrivial task as security is
contextual: a DoS attack on a government network has different behaviors than an
attack on a family business. Therefore, any sort of simulation to observe attack
behavior must be in a similar context to the one in which the dataset was built
from. This is particularly challenging for old datasets like the NSL-KDD, as even
a similar network setup would have different behavior given the modern hardware
and software on the systems that would constitute that network. Even if all these
factors could be accounted for, there are certain classes of attacks whose
``success'' is challenging to measure. For example, in reconnaissance attacks it
is difficult to know if the information gathered by an adversary is useful.

\looseness=-1
Regardless, all of these points of contention are driven by a single hypothesis:
\textit{Perhaps adversarial examples cannot be crafted if features which represent
the semantics of the attack cannot be perturbed}. Instead of attempting to justify
why any set of features is critical to the semantics of the attack, we take a
different stance on addressing this hypothesis: even if some reasonably sized
subset of features could not be perturbed (as to not invalidate the attack), we
argue that adversarial examples can still be successfully crafted. With the
lessons learned through this research, we set out to investigate this hypothesis
with a simple experiment.

\begin{figure}

  \includegraphics[width=\columnwidth]{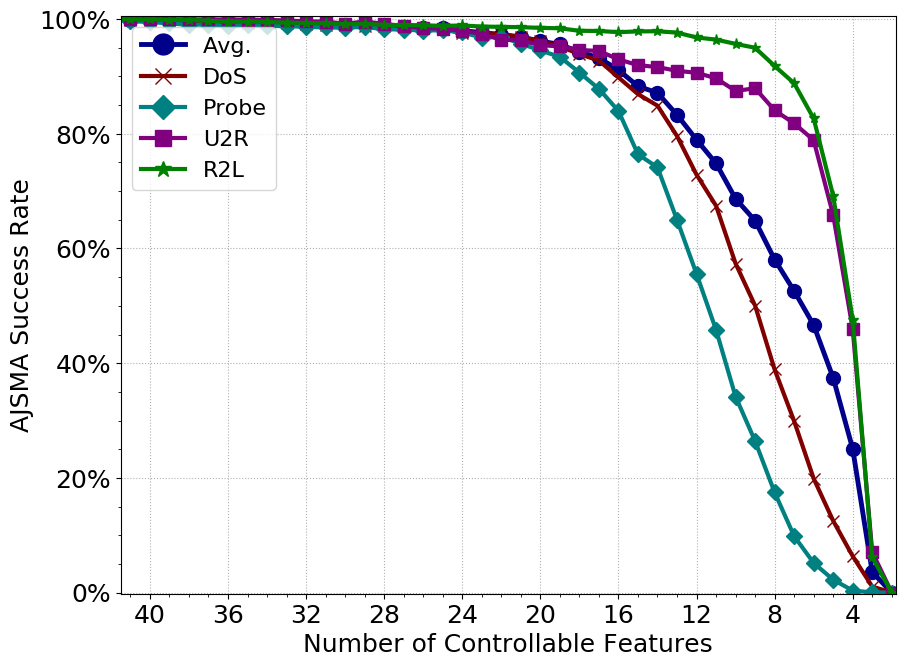}

  \caption{\ajsma{} Success Rate with unperturbable features - The overall success
  rate starts to decrease significantly when the adversary is restricted to controlling 
  $\mathtt{\sim}12$ features.}
  
  \label{fig:fixed}
\end{figure}

\looseness=-1
To test this hypothesis, we iterate over multiple sets of randomly selected
features and make them unperturable by the adversary. We repeat this random
selection for sets of varying cardinalities from 1 to 41 features (where 41
represents the entire feature space). We train a new model on the full NSL-KDD
training set and use the 100 most representative\footnotemark\footnotetext{We find
the most representative inputs by maximizing the difference (via the softmax
layer) between the output component that corresponds to the label and the sum of
the components that represent all non-label classes.} inputs from each
class\footnotemark\footnotetext{The ``R2L'' class only had 17 inputs that were
correctly classified. Thus, we crafted from 317 inputs as opposed to 400.} from
the test set. Next, we iterate over all of the possible combinations of features
made unperturbable to an adversary, i.e., ${\binom{41}{k}}$\footnotemark for
\footnotetext{The NSL-KDD has 41 features before expanding categorical features to
one-hot vectors. If a particular combination contains a categorical feature, we
eliminate all possible values associated with the feature from the search domain.}
$k \in \{1, 2,\dotsc, 41\}$\footnotemark. Once we have identified a set of
unperturable features, we simply eliminate that set from initial search domain of
the \ajsma{}. In this experiment, we crafted a total of 17,664,191 adversarial
examples. Figure \ref{fig:fixed} demonstrates the success rate of the \ajsma{} as
a function of the number of controllable (i.e., perturbable) features.

\footnotetext{To prevent combinatoric explosion, we randomly sampled 1,500 unique
combinations if the total number of possible combinations for a particular value
of $k$ exceeded 1,500. In total, we evaluated 55,723 unique combinations of
unperturbable features.}

\looseness=-1 The results support our argument: the success rate of \ajsma{}
begins to decline when the adversary can only control $\mathtt{\sim}12$ features.
Even when an adversary has control over only \textit{five random} features (which
represents around $\mathtt{\sim}10\%$ of the feature space), the success rate of
crafting adversarial examples (with the most representative forms of an attack) is
slightly less than 50\%. Furthermore, this result demonstrates that applying more
restrictive constraints would have little impact on the success of the adversary. 

Finally, we would also like to highlight how this experiment demonstrates a type
of constraint not covered in the evaluation section: features that the adversary
simply does not have control over. Throughout this work, our constraints were
defined via the semantics of the domain, i.e. the TCP/IP protocol. However, this
experiment to preserve the semantics of the attack also serves as a demonstration
of the efficacy of an adversary under this second type of constraints. These
results suggest that even if an adversary maintains attack behavior \textit{and}
cannot arbitrarily control certain features \textit{and} must obey the TCP/IP
protocol, there is still a surprising amount of exploitable attack surface to
craft legal adversarial examples.

\section{Discussion} \label{sec:discussion} In this section, we describe our
thoughts for future work.

\begin{figure}[t]

  \includegraphics[width=\columnwidth]{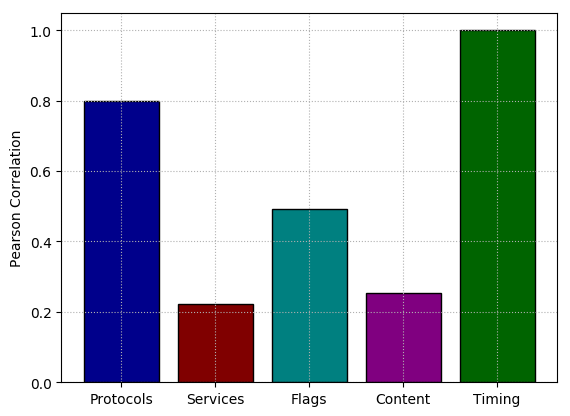}

  \caption{Normalized mean of summed absolute values of Pearson Correlation
  Coefficients for five categories of features in the NSL-KDD - Protocols are
  correlated with the majority of features, closely behind timing-based features,
  which are highly correlated amongst themselves.} \label{fig:correlations}

\end{figure}

\looseness=-1
\vspace{3pt}\noindent\textbf{Identifying Primary Features.} In this work, we
identified primary features manually through not only our understanding of the
domain, but observations of the data. We noticed that many features were
correlated with the transport layer protocol (as we expected, based on the
descriptions of the features). However, we may not always be able to use the
descriptions of features to guide us towards identifying primary features.  Thus,
we hypothesize that primary features could be identified by ranking the features
that are most correlated with others. The intuition here is that while secondary
features will be highly correlated with primary features, primary features will be
highly correlated with the majority of secondary features. We performed a simple
experiment where we computed Pearson product-moment correlation coefficients for
all features in the NSL-KDD training set. We took the mean of sum of the absolute
value of the coefficients for four categories of features as their scores and
found protocols to indeed be correlated with the majority of features, as shown in
Figure \ref{fig:correlations}. At first, it appeared that timing-based features
had a higher correlation than protocols among all features. However, we noticed
that the timing-based features were highly correlated \textit{amongst themselves},
which was the largest contributor to their scores (and to no surprise, since many
timing-based features are derivatives or direct inverses of one another, e.g.,
\textsc{same\_srv\_rate} and \textsc{diff\_srv\_rate}). Thus, if we eliminate such
functionally inverse and derivative features, this approach would suggest that
primary features could be identified systematically.

\looseness=-1
\vspace{3pt}\noindent\textbf{Defenses.} We also hypothesize that constraints
\textit{can} be useful in defending against adversarial examples, even though the
constraints found in the studied datasets were ineffective. We are interested if
constraints in other domains, such as malware and spam, can prove to be effective
against adversarial example generation algorithms. It is intuitive that
maintaining the behavior of malware while simultaneously perturbing binary code
would be a challenging problem. 

In addition, we also observe that our approach for building adversarial sketches
can also be used by a defender to assess model vulnerability. In particular, a
defender could design a simple mechanism (driven by the perturbation histograms)
that reveals universal directions that would make the model vulnerable. The
defender can then use this analysis to detect adversarial examples at deployment.
However, an adversary could circumvent detection by selectively perturbing
features that have less impact. Naturally, this would come at a cost of
introducing additional distortion and control over more features, which may be
impractical for an adversary.

\looseness=-1
Finally, we note that defenses against adversarial examples, including detection
techniques, robust optimization, etc., is an open problem. At present, it
is unclear whether or not a generalizable defense exists against adversarial
examples. 

\section{Conclusions} \label{sec:conclusions} This paper investigated the impact
of adversarial examples in constrained domains through the perspective of
traditional adversarial algorithms and universal adversarial perturbations. In
addition to this investigation in unique domains like network intrusion detection,
we introduced two new algorithms: the \textsc{Adaptive JSMA}, which produces
adversarial examples in constrained domains, and the \textsc{Histogram Sketch
Generation}, which generates adversarial sketches: universal adversarial
perturbations that obey domain constraints. Our work demonstrates how adversaries
can craft permissible adversarial examples in constrained domains.

Through our experiments, we observed how biased distributions coupled with low
dimensionality can have a significant impact on model vulnerability, even in the
presence of constraints. Furthermore, we demonstrated how when defender constrains
an adversary to \textit{five random} features, adversarial examples can still be
crafted with a $\mathtt{\sim}50\%$ success rate.

\looseness=-1 Prior to our work, the impact of adversarial learning has been
largely understood in the context of unconstrained domains. We initially
hypothesized that systems whose domains were constrained would be more resilient
to attack algorithms.  However, our investigation suggests the inverse. Our two
algorithms were able to craft adversarial examples with minimal distortion and
with success rates of greater than 95\% that also transferred to other models at
rates up to 93\%.

Indeed, it is unclear if any domain is immune to adversarial machine learning.
Through a simple number of transformations, an adversary can wholly control a
model, thereby defeating systems deployed in sensitive domains.

\bibliographystyle{ACM-Reference-Format}
\bibliography{bibliography-sheatsley} 
\newpage
\onecolumn
\section{Appendix} \label{sec:appendix}

\begin{table*}[!ht]\footnotesize
  \resizebox{0.76\paperwidth}{!}{ \centering
  \begin{tabulary}{\linewidth}{lLLLL}
    \toprule

    {} & \multicolumn{4}{c}{Feature Type}\\
    \cmidrule(l{1em}r{1em}){2-5}
    {Protocol} & \multicolumn{1}{c}{Basic} & \multicolumn{1}{c}{Content} 
    & \multicolumn{1}{c}{Timing-based} & \multicolumn{1}{c}{Host-basted}\\
    \midrule

    {\textbf{TCP}} & {duration, service=aol, service=auth, service=bgp,
    service=courier, service=csnet\_ns, service=ctf, service=daytime,
    service=discard, service=domain, service=echo, service=efs, service=exec,
    service=finger, service=ftp, service=ftp\_data, service=gopher,
    service=harvest, service=hostnames, service=http, service=http\_2784,
    service=http\_443, service=http\_8001, service=IRC, service=iso\_tsap,
    service=klogin, service=kshell, service=ldap, service=link, service=login,
    service=mtp, service=name, service=netbios\_dgm, service=netbios\_ns,
    service=netbios\_ssn, service=netstat, service=nnsp, service=nntp,
    service=other, service=pm\_dump, service=pop\_2, service=pop\_3,
    service=printer, service=private, service=remote\_job, service=rje,
    service=shell, service=smtp, service=sql\_net, service=ssh, service=sunrpc,
    service=supdup, service=systat, service=telnet, service=time, service=uucp,
    service=uucp\_path, service=vmnet, service=whois, service=X11,
    service=Z39\_50, service=34, flag=OTH, flag=REJ, flag=RSTO, flag=RSTOS0,
    flag=RSTR, flag=S0, flag=S1, flag=S2, flag=S3, flag=SF, flag=SH, src\_bytes,
    dst\_bytes, land=1, urgent} & 

    {hot, num\_failed\_logins, logged\_in=1, num\_compromised, root\_shell,
    su\_attempted, num\_root, num\_file\_creations, num\_shells,
    num\_access\_files, is\_host\_login=1, is\_guest\_login=1} & 
   
    {count, srv\_count, serror\_rate, srv\_serror\_rate, rerror\_rate,
    srv\_rerror\_rate, same\_srv\_rate, diff\_srv\_rate, srv\_diff\_host\_rate} & 

    {dst\_host\_count, dst\_host\_srv\_count, dst\_host\_same\_srv\_rate,
    dst\_host\_diff\_srv\_rate, dst\_host\_same\_src\_port\_rate,
    dst\_host\_srv\_diff\_host\_rate, dst\_host\_serror\_rate,
    dst\_host\_srv\_serror\_rate, dst\_host\_rerror\_rate,
    dst\_host\_srv\_rerror\_rate} \\

    \midrule

    {\textbf{UDP}} & {duration, service=domain\_u, service=ntp\_u, service=other,
    service=private, service=tftp\_u, flag=SF, wrong\_fragment} & 

    {} & 

    {count, srv\_count, serror\_rate, rerror\_rate, same\_srv\_rate,
    diff\_srv\_rate, srv\_diff\_host\_rate} & 

    {dst\_host\_count, dst\_host\_srv\_count, dst\_host\_same\_srv\_rate,
    dst\_host\_diff\_srv\_rate, dst\_host\_same\_src\_port\_rate,
    dst\_host\_srv\_diff\_host\_rate, dst\_host\_serror\_rate,
    dst\_host\_rerror\_rate} \\

    \midrule

    {\textbf{ICMP}} & {service=eco\_i, service=ecr\_i, service=red\_i, service=tim\_i,
    service=urh\_i, service=urp\_i, flag=SF, src\_bytes, wrong\_fragment} & 

    {} & 

    {count, srv\_count, serror\_rate, rerror\_rate, same\_srv\_rate,
    diff\_srv\_rate, srv\_diff\_host\_rate} & 

    {dst\_host\_count, dst\_host\_srv\_count, dst\_host\_same\_srv\_rate,
    dst\_host\_diff\_srv\_rate, dst\_host\_same\_src\_port\_rate,
    dst\_host\_srv\_diff\_host\_rate, dst\_host\_serror\_rate,
    dst\_host\_rerror\_rate} \\

    \bottomrule
 \end{tabulary}}

  \caption{The constraints extracted from the NSL-KDD - Unlike TCP, UDP and ICMP
  have limited degrees of freedom.} \label{tab:nsl-constraints}

\end{table*}

\begin{table*}[!ht]\footnotesize
  \resizebox{0.55\paperwidth}{!}{ \centering
  \begin{tabular}{rrrcccc}
    \toprule
  
    {Learning Technique} & \multicolumn{2}{c}{Parameters} & \multicolumn{4}{c}{Test Accuracy} \\
    \cmidrule(l{0.5em}r){2-3} \cmidrule(l{0.5em}r{0.5em}){4-7}
    {} & \multicolumn{1}{c}{Name} & \multicolumn{1}{c}{Value} & {NSL-KDD} & {UNSW-NB15} & {MNIST} & {GTSRB} \\
    \midrule
    {\textbf{Logistic Regression}} & 
      \begin{tabular}{@{}r@{}}\textsc{penalty}\\\textsc{C}\end{tabular} &
        \begin{tabular}{@{}r@{}}l2\\1.0\end{tabular} & {74.21\%} & {67.65\%} & {92.02\%} & {84.94\%}\\
    \cmidrule(l{0.5em}r){2-3} \cmidrule(l{0.1em}){4-7}
    {\textbf{Support Vector Machine}} & 
      \begin{tabular}{@{}r@{}}\textsc{C}\\\textsc{kernel}\\\textsc{degree}\end{tabular} & 
        \begin{tabular}{@{}r@{}}1.0\\rbf\\3\end{tabular} & {77.31\%} & {69.02\%} & {94.04\%} & {84.3\%}\\
    \cmidrule(l{0.5em}r){2-3} \cmidrule(l{0.1em}){4-7}
    {\textbf{Decision Tree Classifier}} &
      \begin{tabular}{@{}r@{}}\textsc{criterion}\\\textsc{max\_depth}\\\textsc{min\_samples\_split}\\
      \textsc{min\_samples\_leaf}\\\textsc{max\_features}\end{tabular} & 
        \begin{tabular}{@{}r@{}}gini\\$\infty$\\2\\1\\$\infty$\end{tabular} & {74.52\%} & {73.27\%} & {87.73\%} & {57.20\%}\\
    \cmidrule(l{0.5em}r){2-3} \cmidrule(l{0.1em}){4-7}
    {\textbf{k-Nearest Neighbor}} &
      \begin{tabular}{@{}r@{}}\textsc{k}\\\textsc{p}\\\end{tabular} &
        \begin{tabular}{@{}r@{}}5\\2\end{tabular} & {74.90\%} & {72.20\%} & {96.88\%} & {52.83\%}\\

    \bottomrule
  \end{tabular}}
  \caption{Scikit-Learn Model Information}
  \label{tab:scikit-architecture}

\end{table*}

\begin{figure*}[!ht]
  \begin{subtable}[h]{0.5\paperwidth}
      \includegraphics[width=\textwidth]{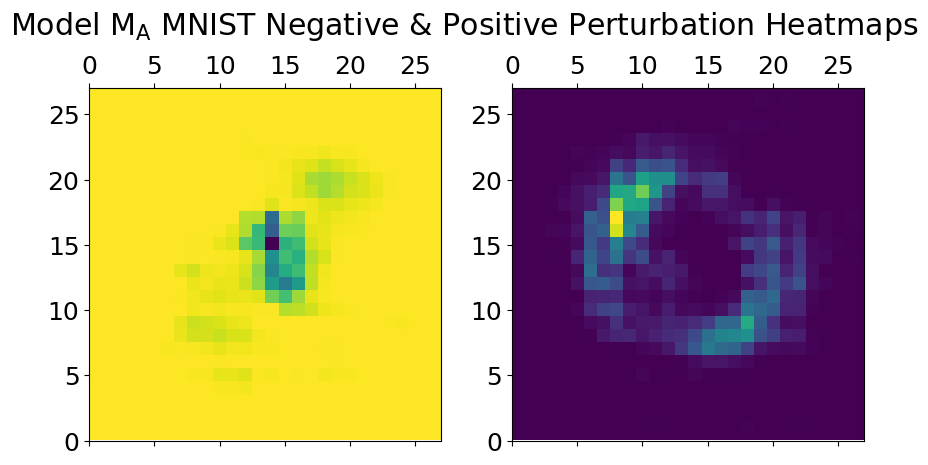}
      \caption{Perturbation Histogram produced by the AJSMA for the MNIST model $M_A$
      as a 2D Heatmap - The target class (``0'') can be visibly seen by combining the
      decreasing features plot (left) with the increasing features plot (right).}
  \end{subtable}
  \begin{subtable}[h]{0.52\paperwidth}
      \includegraphics[width=\textwidth]{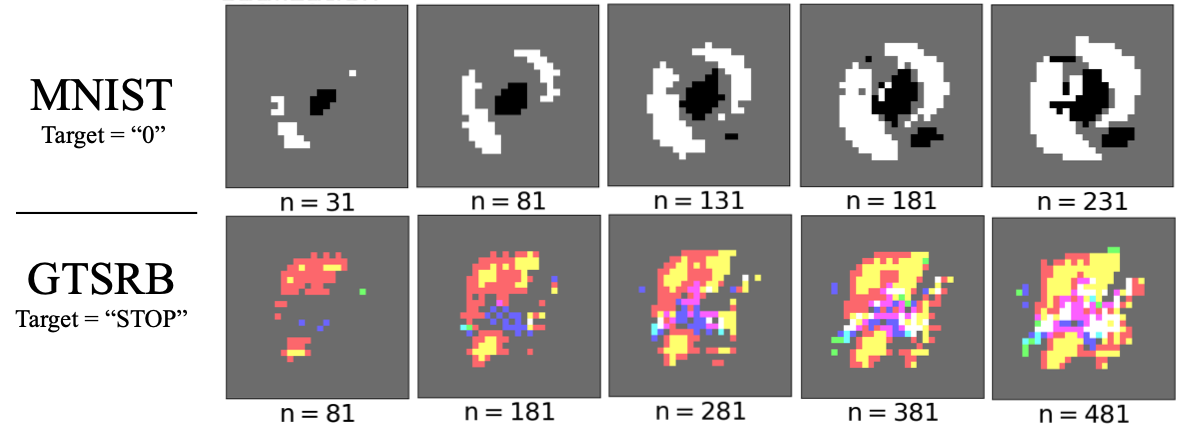}
      \caption{Sketch Visualization for MNIST and GTSRB for varying values of $n$ - 
      As $n$ increases, the target class visibly forms.}
  \end{subtable}
  \caption{Perturbation Histogram for MNIST (a) and adversarial sketches for image
  datasets (b).}
  \label{fig:graphics}
\end{figure*}

\end{document}